\documentclass[12pt]{article}
\usepackage{a4}
\usepackage{epsfig}

\makeatletter
\@addtoreset{equation}{section}
\makeatother

\begin{document}

\newcommand{\bee}{\begin{equation}}
\newcommand{\eeq}{\end{equation}}
\newcommand{\bea}{\begin{eqnarray}}
\newcommand{\eea}{\end{eqnarray}}
\newcommand{\pa}{\partial}
\newcommand{\tl}{\tilde{\lambda}}
\newcommand{\tg}{\tilde{g}}
\newcommand{\MSb}{{\overline {\rm MS}}}
\newcommand{\df}{\delta f(h)}
\newcommand{\eff}{\lambda_{\scriptscriptstyle {\rm eff}}}
\newcommand{\lb}{\lbrack}
\newcommand{\Se}{S_{\rm eff}}
\newcommand{\rb}{\rbrack}

\begin{titlepage}
\title{A 2D integrable axion model and Target space duality
\vspace{2.5truecm} }
\date{}
\author{
{\Large P\'{e}ter Forg\'{a}cs}\\
 \\
{\small Laboratoire de Math\'{e}matiques et Physique Th\'{e}orique}\\
{\small CNRS UMR 6083}\\
{\small Universit\'{e} de Tours}\\
{\small Parc de Grandmont, 37200 Tours, France}\\
 \vspace{1truecm}
 }

\maketitle

\abstract{A review is given on the recently
proposed two dimensional axion model
(O(3) $\sigma$-model with a dynamical $\theta$-term)
and the T-duality relating it to the
SU(2)$\times$U(1) symmetric anisotropic $\sigma$-model.
Strong evidence is presented
for the correctness of the proposed $S$-matrix for both models
comparing perturbative and Thermodynamical
Bethe Ansatz calculations for different types of free energies.
This also provides a very stringent test of the validity of T-duality
transformation at the quantum level.
The quantum non-integrability of
the O(3) $\sigma$-model with a {\sl constant} $\theta$-term,
in contradistinction to the axion model,
is illustrated by calculating
the $2\rightarrow3$ particle production amplitude
to lowest order in $\theta$.
}

\end{titlepage}

\setcounter{section}{0} \setcounter{equation}{0}
\section{The axion model}

Let us consider
the following two dimensional
$\sigma$-model defined by the Lagrangian\footnote{The following conventions are
used: for
a vector $v$ in two-dimensional Minkowski space
$v^\mu v_\mu=v_+v_-$ where $v_\pm=v_0\pm v_1$.
The antisymmetric tensor is defined by $\epsilon^{01}=1$,
$\tau^a=\sigma^a/2$
with $\sigma^a$ being the standard Pauli-matrices satisfying
$\sigma^a\sigma^b=\delta^{ab}+i\epsilon^{abc}\sigma^c$,
$a,b,c=1,2,3$ and summation is implied over the repeated
indices.}:
\bee
{\cal L}=
{1\over2\tilde\lambda}\partial_\mu n^a \partial^\mu n^a+
{\tilde\lambda\over32\pi^2(1+\tilde g)}
\partial_\mu\theta\partial^\mu\theta+
{\theta\over8\pi}\epsilon^{\mu\nu}\epsilon^{abc}
n^a\partial_\mu n^b\partial_\nu n^c\,,
\label{Lagax}
\eeq
where $\tilde\lambda$, $\tilde g$ are (real) parameters
(couplings) and the $n$-fields are subject to the
constraint $n^an^a=1$. I shall refer to (\ref{Lagax}) as the
`axion model' since it is the O(3) nonlinear $\sigma$-model with
a {\sl dynamical} $\theta$-term, which is in turn a natural
two-dimensional analogue of its phenomenologically important
four-dimensional counterpart, in the same way as the
O(3) $\sigma$-model is considered to be a 2D analogue of 4D
non-abelian gauge theories. As will be exhibited later
(\ref{Lagax}) is asymptotically free for the coupling range
$-1<\tilde g\le0$. The theory defined by
(\ref{Lagax}) is an O(3) non-linear $\sigma$-model,
coupled to a scalar field, $\theta$, (whose normalization has
been chosen for later convenience)
through the Hopf term. The latter is proportional
to the topological current of the O(3) model and with the
normalization chosen in (\ref{Lagax}) its space-time integral
(after a Wick rotation) yields the topological charge, which is
integer valued (for non singular configurations). 
The variable $\theta$ therefore has a natural interpretation as an angle,
$\theta\in[0,2\pi]$, so that the three dimensional
target space of the axion model is topologically $S^2\times S^1$.
Although at the classical level only one of the parameters
(couplings), ($\tilde\lambda$, $\tilde g$), is relevant I prefer to
introduce them from the very beginning as
in the quantum theory both couplings play a r\^ole. It is not
difficult to see that for $\tilde g\to -1$ the $\theta$ field
decouples from $n^a$, i.e.\ one obtains the O(3) $\sigma$-model and a
decoupled free scalar field. There is another special value of the
coupling,
$\tilde g=0$, when (\ref{Lagax}) turns out to be canonically equivalent
to the O(4) $\sigma$-model.

The `manifest' global symmetries of the axion model are
SO(3)$\times$SO(2)$_\theta$, where SO(2)$_\theta$ generates a
shift $\theta\to\theta+{\rm const.}$, a symmetry of
(\ref{Lagax}) up to a total derivative.

In this paper I shall review the axion model and its
dual, the SU(2)$\times$U(1) symmetric anisotropic principal $\sigma$-model.
The most remarkable feature of the axion model
is its conjectured quantum integrability (in the sense that there is no particle
production). It allows one to deduce by bootstrap methods
its exact spectrum, which turns out to depend on the value of the
(renormalization group invariant) coupling ratio
$\tilde p={2\pi(1+\tilde g)/\tilde\lambda}$.
As long as $0<\tilde p<1$ the spectrum consists of SO(3)
triplets and singlets (`breathers') as well as of doublets
(`kinks'), while for $\tilde p>1$ only two doublets.

This review is organized in the following way:
In Section 2 it is shown
that the classical equations of motion
of the axion model admit a (standard) Lax pair,
implying integrability (in the sense that an infinite set conserved
quantities exists).
Next in Subsection 2.2 the axion model is demonstrated
to be equivalent to the rather well
studied SU(2)$\times$U(1) symmetric anisotropic $\sigma$-model
by a simple target-space duality transformation.
This observation leads to a new Lax pair.\\
Section 3 is devoted to perturbative investigations of the
two models.
First a brief renormalization group analysis of the anisotropic
model is carried out, emphasizing the important property of
asymptotic freedom.
Then the two loop $\beta$-functions are compared
in the two models with the conclusion that they are
equivalent provided one takes into account a change of the renormalization
scheme.
In Subsections 3.2, and 3.3 the Legendre transformation of
various (bosonic and fermionic type) finite density ground state energies
(free energies) is computed in the anisotropic model.
In Subsection 3.4 the main points of the analogous computation in the axion
model are discussed. It is found that the free energies
fully agree in perturbation theory.
Since one compares physical quantities in the two (dual) models,
no scheme dependence arises and therefore this provides good evidence
for the validity of a quantum version of this T-duality.\\
The anisotropic $\sigma$-model is generally believed to be quantum
integrable.
Applying the thermodynamical Bethe Ansatz
one can also compute the free energies nonperturbatively from the
bootstrap
$S$-matrix and the comparison with the perturbative results
yields quite a stringent test of the proposed $S$-matrix.
This program is carried out in Section 4.\\
First in Subsection 4.1, in the framework of the form factor bootstrap
approach, the $2\to3$ particle production amplitude
in the O(3) $\sigma$-model with a {\sl constant} $\theta$-term
is shown to be non vanishing.
To the best of my knowledge this is the first {\sl quantitative} evidence
for the (generally expected)
quantum non-integrability of this model.
To my mind this fact underlines the rather surprising nature
of the quantum integrability of the axion model.
Subsection 4.2 is devoted to the thermodynamical Bethe Ansatz and in
4.3 the exact relation
between the infrared and the ultraviolet mass scales ($m/\Lambda$-ratio)
is given. Finally in Subsection 4.4 the limit $\tilde p\to0$
(O(3) limit) is discussed, in particular it is explained how one recovers
the known $m/\Lambda$-ratio of the O(3) $\sigma$-model.

\setcounter{section}{1} \setcounter{equation}{0}
\section{Lax pairs and T duality}

 \subsection{A Lax pair for the axion model}

The axion model belongs to a family of O(3) symmetric {\sl
classically integrable} $\sigma$-models discovered in
Ref.\ \cite{class}, some of whose results I now recall.
Introducing the matrix valued current
\bee
I_\mu={\tilde\lambda\over8\pi}n\epsilon_{\mu\nu}\partial^\nu\theta-
{\sqrt{\tilde g}\over2}\epsilon_{\mu\nu}\partial^\nu n+{1\over2}
n\partial_\mu n\,,
\label{I}
\eeq
where $n=in^a\sigma^a$, the
equations of motion of (\ref{Lagax}) can be written as:
\bee
\partial^\mu I_\mu=0\,,\qquad\qquad
\partial_\mu I_\nu-\partial_\nu I_\mu=[I_\mu,I_\nu]\,.
\label{LaxI}
\eeq
The standard form (\ref{LaxI}) of the equations of motion
allows for the introduction of a Lax pair,
\bee
U_\pm={1\over1\pm\omega}\,I_\pm\,,
\label{LaxU}\eeq
satisfying the zero curvature equation
\bee
\partial_\mu U_\nu-\partial_\nu U_\mu=[U_\mu,U_\nu]\,,
\label{zeroU}
\eeq
for all values of the spectral parameter $\omega$.
The current, $I_\mu$, is closely related to the matrix valued
Noether current, ${\cal N}_\mu=-i\tau^a{\cal N}^a_\mu$,
defined by
$\delta{\cal L}=\partial^\mu\varepsilon^a{\cal N}^a_\mu$
corresponding to the symmetry
transformation $\delta n^a=\epsilon^{abc}\varepsilon^bn^c$:
\bee
I_\mu=\tilde\lambda{\cal N}_\mu+\epsilon_{\mu\nu}\partial^\nu T\,,
\qquad\qquad T=\Big({\tilde\lambda\over8\pi}\theta-{\sqrt{\tilde g}
\over2}\Big)n\,.
\label{IN}
\eeq
Note that the trivially conserved part of $I_\mu$
($\epsilon_{\mu\nu}\partial^\nu T$) is essential for
the zero curvature equation (\ref{zeroU}) to be satisfied.

\subsection{T-dual of the axion model}

Here I show that the axion model
(\ref{Lagax}) is classically equivalent
to a rather well studied $\sigma$-model,
the so-called anisotropic SU(2) principal $\sigma$-model (PCM).
The equivalence is a canonical one given by an Abelian target space 
(T-)duality transformation \cite{Busch,Alvarez}).

The Lagrangian of the anisotropic PCM can be written as:
\bee
{\cal L}_\Sigma=-{1\over2\lambda}\Big\{L^a_\mu\,L^{a\mu}+
gL^3_\mu\,L^{3\mu}\Big\}\,,
\label{Lagdef}
\eeq
where
\bee
L_\mu=G^{-1}\partial_\mu G=\tau^aL^a_\mu\,,
\label{L}
\eeq
and $g$ is the parameter of anisotropy.
The Lagrangian (\ref{Lagdef}) can be interpreted as a deformation of
the SU(2)$\times$SU(2) (or O(4)) symmetric nonlinear $\sigma$-model
by the parameter $g$. For a generic value of
$g$ this (torsionless) model has
an SU(2)$_{\rm L}\times$U(1)$_{\rm R}$ symmetry.
At the classical level the anisotropic  model, (\ref{Lagdef}),
interpolates
between the SU(2)$\times$SU(2) ($g=0$) and O(3) ($g=-1$) models.

The equations of motion of the anisotropic principal model
(\ref{Lagdef})
can be written entirely in terms of the current $L_\mu$:
\bee
\partial^\mu L^3_\mu=0\,,\quad
\partial^\mu L^1_\mu=-igL^{2\mu}L^3_\mu\,,\quad
\partial^\mu L^2_\mu=igL^{1\mu}L^3_\mu\,.
\label{EOM}
\eeq
It is known that the equations of motion (\ref{EOM}) admit a Lax
representation
\cite{Cherednik,Kirillov}, i.e.\ there is a spectral parameter dependent
current, $V_\mu=\tau^aV^a_\mu$, satisfying the zero
curvature equation (\ref{zeroU}). This current can be written as:
\bee
V^{1,2}_\pm=\alpha_\pm\,L^{1,2}_\pm\,,\qquad\qquad
V^3_\pm=a_\pm\,L^3_\pm\,,
\label{V}
\eeq
where
\bee
\alpha_\pm=-{4+g\omega^2\over4-g\omega^2\pm4\omega}\,,\qquad
a_\pm=-{4-g\omega^2\mp4g\omega\over4-g\omega^2\pm4\omega}\,.
\label{alphapm}
\eeq

To exhibit the classical T-duality transformation
mapping the axion model (\ref{Lagax}) to the anisotropic PCM
(\ref{Lagdef}) \cite{bfhp}, it is convenient to parametrize the $n^a$-fields in
Eq.\ (\ref{Lagax}) as
\bee
n^1=\sin\vartheta\sin\varphi\,,\quad
n^2=\sin\vartheta\cos\varphi\,,\quad
n^3=\cos\vartheta\,,\qquad
\theta=-{4\pi\over\tilde\lambda}\sqrt{1+\tilde g}\,\chi\,,
\label{para_ax}
\eeq
in terms of which the Lagrangian of the axion model (\ref{Lagax}) (after
an integration by parts) becomes
\bee
{\cal L}={1\over2\tilde\lambda}\Big\{
\partial_\mu\vartheta\partial^\mu\vartheta
+\sin^2\vartheta\partial_\mu\varphi\partial^\mu\varphi
+\partial_\mu\chi\partial^\mu\chi
+2\sqrt{1+\tilde g}\cos\vartheta\epsilon^{\mu\nu}
\partial_\mu\chi\partial_\nu\varphi\Big\}\,.
\label{Lagax_par}
\eeq
An Abelian T-duality transformation \cite{Busch} with respect to the
SO$(2)_{\theta}$ symmetry of the axion model (\ref{Lagax}), corresponds
to the following canonical transformation \cite{Alvarez}:
\bee
\chi^\prime=-{\tilde\lambda\over\sqrt{1+\tilde g}}p_\alpha\qquad\qquad
p_\chi=-{\sqrt{1+\tilde g}\over\tilde\lambda}\alpha^\prime\,,
\label{cano}
\eeq
where (and in the following) $p_\chi$ resp.\ $p_\alpha$ denote the
canonical momenta conjugate to $\chi$ resp.\ to its `dual'
$\alpha$. In terms of the new variable, $\alpha$,
the dual Lagrangian turns out to be:
\bee
{\cal L}_\Sigma={1\over2\tilde\lambda}\Big\{
\partial_\mu\vartheta\partial^\mu\vartheta
+(1+\tilde g\cos^2\vartheta)\partial_\mu\varphi\partial^\mu\varphi
+(1+\tilde g)\left[\partial_\mu\alpha\partial^\mu\alpha
+2\cos\vartheta
\partial_\mu\alpha\partial^\mu\varphi\right]\Big\}\,.
\label{Lagdef_par}
\eeq
A simple calculation shows that (\ref{Lagdef_par}) is nothing but
the Lagrangian (\ref{Lagdef}), when parametrizing the SU(2) valued
field, $G$, by the Euler angles
\bee
G=e^{i\varphi\tau^3}\,e^{i\vartheta\tau^1}\,
e^{i\alpha\tau^3}\,,
\label{G}
\eeq
and taking into account the relations at the classical level
between the couplings:
\bee\label{couplrel}
\tilde\lambda =\lambda\,,\quad
\quad \tilde g=g\,.
\eeq
The observation that the axionic model is the T dual of
${\cal L}_\Sigma$ also explains why $\theta$ should be interpreted
as an angular variable.
Indeed, as it has been shown in Ref.\ \cite{alvtop}
the Abelian T duality transformation
(\ref{cano}) maps the target space of the isotropic PCM, $S^3$,
the Abelian T duality (\ref{cano}) into  $S^2\times S^1$.
The arguments of
Ref.\ \cite{alvtop} can be easily adopted to the present case with
$g>-1$, and it is clear that in Eq.\ (\ref{Lagax}) $n^a$ parametrize an
$S^2$ and $\theta $ parametrizes an $S^1$.

One can now use the classical T-duality transformation
(\ref{cano}) to map the linear system of the axion model
(\ref{LaxI}) to a new Lax pair for the anisotropic $\sigma$-model
(\ref{Lagdef}). It is given by Eq.\ (\ref{LaxU}), where the
current, $I_\mu$, has to be replaced by
\bee
\hat I_\mu=\partial_\mu G\,G^{-1}+{g} \Big(G\tau^3G^{-1}\Big)
L^3_\mu-{i\sqrt{g}}\epsilon_{\mu\nu}\partial^\nu
\Big(G\tau^3G^{-1}\Big)\,.
\label{dualI}
\eeq
Eq.\ (\ref{dualI})
is obtained from (\ref{I}) by the T-duality transformation
(\ref{cano}). $\hat I_\mu$ is related to the Noether current
$\hat{\cal N}_\mu$, corresponding to the manifest symmetry
$\delta G=-i\varepsilon^a\tau^a G$ of (\ref{Lagdef}) and can be
written analogously to ${\cal N}_\mu$:
\bee
\hat I_\mu=\lambda\hat{\cal N}_\mu+ \epsilon_{\mu\nu}\partial^\nu
 \hat T\,, \qquad\qquad \hat T=-{i\sqrt{g}} \, G\tau^3G^{-1}\,.
\label{IhatN}
 \eeq
It is clear that the new Lax pair (\ref{LaxU})
and the \lq old' one, (\ref{V}), cannot be related by a gauge
transformation since they have different pole structures as
functions of the spectral variable, $\omega$. In the $\tilde g\to-1$
limit, the axion model reduces to the original O(3)
$\sigma$-model (decoupled from the $\theta$ field), and the Lax
pair (\ref{LaxU}) becomes equivalent to that of Ref.\ \cite{Byt},
where it has been pointed out that the corresponding $\hat
I_\mu$'s are {\sl ultralocal} currents. I remark that the Lax
pairs (\ref{LaxU}) and (\ref{V}) correspond to (different)
 deformations
of the usual Lax pairs of the principal chiral $\sigma$-model,
linear in $\partial_\mu GG^{-1}$ respectively $G^{-1}\partial_\mu G$.

\setcounter{section}{2} \setcounter{equation}{0}
\section{Quantum equivalence}

\subsection{Perturbative investigations}

The canonical equivalence between the axion and the anisotropic PCM
models naturally raises the question if there is also a quantum
equivalence between them. Or putting it a slightly different way, one
can
ask if there is a quantum version of T-duality transformation (\ref{cano}).

The equivalence of the models in perturbation theory (PT)
is certainly a necessary
condition for the validity of quantum T-duality.
I start by recalling some renormalization properties of the anisotropic
$\sigma$-model and the important property of asymptotic freedom.
Then I present a quantitative comparison in PT between the two models
by comparing their $\beta$-functions.

Due to its SU(2)$_{\rm L}\times$U(1)$_{\rm R}$ symmetry, the anisotropic
$\sigma$-model (\ref{Lagdef})
is renormalizable in perturbation theory. (Of course, in Eq.\
(\ref{Lagdef})
$(\lambda,g)$ should be interpreted as the bare couplings
$(\lambda_0,g_0)$.)
Using dimensional regularization, the relation between the
bare parameters, $(\lambda_0,g_0)$, and the renormalized ones,
$(\lambda,g)$, is written as:
\bee
\lambda_0=\mu^\epsilon\lambda\,Z_\lambda\,,\qquad\qquad{\rm and}
\qquad\qquad 1+g_0=(1+g)\,Z_g\,.
\label{bare}
\eeq
In Eq.\ (\ref{bare}) the dimensional parameter, $\mu$, is introduced as
usual, to carry the mass dimension of
$\lambda_0$ (which is dimensionless in 2D).
The renormalization constants contain only simple poles in
$\epsilon=2-n$ whose
residues can be calculated in perturbation theory.

Physical quantities depend on the renormalized couplings $\lambda$,
$g$ and the dimensional parameter, $\mu$, in such a way that the action
of the renormalization group (RG) operator
\bee
{\cal D}=\mu\,{\partial\over\partial\mu}
+\beta_\lambda(\lambda,g)\,{\partial\over\partial\lambda}
+\beta_g(\lambda,g)\,{\partial\over\partial g}\,,
\label{RGop}
\eeq
vanishes on them, where the $\beta$-functions are obtained from the
residues of the first order poles.
The RG equations
for the running couplings $\bar\lambda$, $\bar g$ can be written as
\bea
{d\bar\lambda\over dt} =\beta_\lambda(\bar\lambda,\bar g)\,,&
\qquad\qquad
&\bar\lambda(0)=\lambda\,,\label{RGeq1}\\
{d\bar g\over dt} =\beta_g(\bar\lambda,\bar g)\,,&
\qquad\qquad\,
&\bar g(0)=g\,.\label{RGeq2}
\nonumber
\eea
One is interested in the `large time' asymptotic
behaviour of the couplings as functions of an external field
(or chemical potential)
$h=h_0\,e^t$, with $h_0$ fixed and $t\rightarrow\infty$.
In the anisotropic PCM the $\beta$-functions are known to be
of the form:
\bea
\beta_\lambda&=& -{\lambda^2\over4\pi}\Big\{1-g+\lambda p_2(g)+
\cdots\Big\}\,,\label{beta1}\\
\beta_g&=& {\lambda g(1+g)\over2\pi}\Big\{1+\lambda q_1(g)+
\cdots\Big\}\,,\label{beta2}
\nonumber
\eea
where the two-loop $\beta$-function coefficients are:
\bee
p_2(g)={1-2g+5g^2\over8\pi}\,,\qquad\qquad{\rm and}
\qquad\qquad q_1(g)={1-g\over4\pi}\,.
\label{2loop}
\eeq

Asymptotically free
(AF) behaviour occurs when $\bar\lambda
\rightarrow0$ together with $\bar g\rightarrow g_1$ as
$t\rightarrow\infty$, where
$g_1$ is some constant.
Analyzing Eqs.\ (\ref{beta1}), (\ref{beta2}), one finds
three different AF solutions:
\bea
{\rm 1.}\qquad &g_1=\phantom{-}0&
\qquad\qquad g\equiv g_0\equiv\bar g\equiv0\,,
\label{AF1}\\
{\rm 2.}\qquad &g_1=-1&\qquad\qquad g\equiv g_0\equiv\bar g\equiv -1\,,
\label{AF2}\\
{\rm 3.}\qquad &g_1=-1&\qquad\qquad -1<\bar g<0\,.
\label{AF3}
\eea
Solutions (\ref{AF1}) resp.\ (\ref{AF2}) correspond to the O(4) resp.\
the O(3) $\sigma$-model, while the solution (\ref{AF3}) corresponds to
the \lq generic' case of the anisotropic PCM.

\begin{figure}[t]
\hspace{3.5cm}
\leavevmode
\epsfxsize=15pc
 \epsfbox{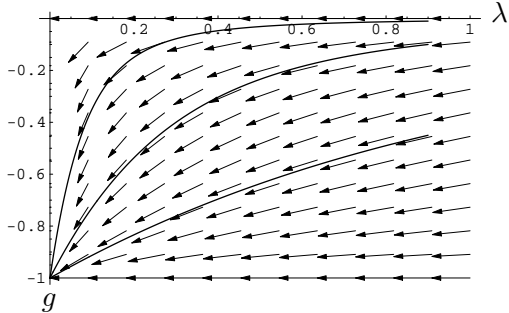}
 \begin{picture}(10,80)\put(-170,-4.0){\small $g$}
 \put(0,103.5){\small$\lambda$}
\end{picture}

\caption{1 loop flow diagram of the couplings $(\lambda,g)$
\label{fig:radish}}
\end{figure}
It is very convenient to introduce an exact RG invariant combination of
the two couplings
\bee\label{invp}
p=2\pi\lim_{t\rightarrow\infty}{1+g(t)\over\lambda(t)}\,,
\eeq
and an effective $\beta$-function for
$\lambda(t)$ by
\bee
\beta_{\scriptscriptstyle{\rm eff}}(\lambda,p)
=\beta_\lambda(\lambda,\Gamma(\lambda,p))\,,
\eeq
expressing
$g(t)$, in terms of the running coupling, $\lambda(t)$,
and the RG invariant quantity $p$ as
$g(t)=\Gamma(\lambda(t)\,,p)$.
Using the perturbative result for $\Gamma(\lambda(t),p)$ \cite{BF}
one finds
\bee
\beta_{\scriptscriptstyle{\rm eff}}(\lambda,p)=
\beta_{\scriptscriptstyle{\rm eff}}(\tilde\lambda,\tilde p)
=-{\lambda^2\over2\pi}+
{p-2\over8\pi^2}\lambda^3+\cdots\,.
\label{betaeff1}
\eeq
Eq.\ (\ref{betaeff1}) allows for the introduction of an RG-invariant
$\Lambda$-parameter in the $\overline{\rm MS}$ scheme in the usual way:
\bee
\Lambda_{\overline{\rm MS}}=\mu\, e^{-{2\pi\over\lambda}}
\Big({\lambda\over
2\pi}\Big)^{\big({p\over2}-1\big)}
\,e^\gamma\Big\{1+{\cal O}(\lambda)\Big\}\,.
\label{Lambda}
\eeq
The effective
coupling $\eff(h)$ is defined by the transcendental equation
\bee
{2\pi\over \eff}+\Big({p\over2}-1\Big)\ln{2\pi\over \eff}=\ln{h\over
\Lambda_{\overline{\rm MS}}}\,,
\label{eff}
\eeq
which is a function of the `physical' parameter
\bee
s=\ln{h\over\Lambda_{\overline{\rm MS}}}\,,
\label{s}
\eeq
only.
Moreover the running coupling can be expressed in terms of
$\eff(h)$ perturbatively (in the sense that it is an infinite power
series).
The asymptotic expansion (for large $s$) of the
effective coupling (containing terms $\propto\ln s$) can be written as
\bee
\eff={2\pi\over s}+{\pi(p-2)\over s^2}\ln s+\cdots\;.
\label{e1}
\eeq

After this detailed discussion of the renormalization properties of the
anisotropic PCM model I just briefly recall the main point for the
axion model.
As discussed in detail in Ref.\ \cite{bfhp} the axion model is also
renormalizable in the field theoretical sense and
at one loop order the $\beta$-functions of
the couplings of the axion model,
$\beta_{\tilde{\lambda}}$, $\beta_{\tilde{g}}$, are simply
obtained from $\beta_{\lambda}$, $\beta_{g}$
by the classical relation (\ref{couplrel}).
At two loops, however, it has been found in \cite{bfhp}
when using the background field method and dimensional regularization
that the following perturbative redefinition of the couplings
\bee\label{psicoupl}
\tilde\lambda=\lambda+{\lambda^2\over4\pi}(1+g)\,,\quad
\tilde g=g+{\lambda\over4\pi}(1+g)^2\,,
\eeq
(i.e.\ a change of scheme)
is induced by the T-duality transformation. Taking into account
Eqs.\ (\ref{psicoupl}) the two loop
$\beta$-functions of the two models turn out to be equivalent.
Alternatively, defining the renormalization group
invariant combination of the two couplings analogously as in the PCM
model Eq.\ (\ref{invp}):
\bee
\tilde p=2\pi\lim_{t\rightarrow\infty}
{1+\tilde g(t)\over\tilde\lambda(t)}\,,
\label{g}
\eeq
one finds
$$p=\tilde p$$
up to two loops \cite{bfhp}.
Thus as far as coupling constant renormalization is concerned,
the two models are equivalent, both
are asymptotically free, and the actual value of $p$ effects only the
two loop coefficient.

\subsection{Perturbative tests of free energies}

Up to now quantum equivalence between the two dually related
models has been tested by comparing simple renormalization properties
and the (two loop) $\beta$-functions.
The fact that the higher coefficients of the $\beta$-functions
are scheme dependent makes such a comparison more
difficult and less conclusive.
To lend some more credibility to the proposed perturbative quantum
equivalence between the two models it is clearly desirable to compare
{\sl physical quantities}.
For this purpose
it is quite convenient to compute various zero temperature free energies
obtained
by minimizing the Legendre transform of the Hamiltonian density
coupled to some conserved currents
\bee\label{hamgen}
\hat{\cal H}={\cal H}_0-h_iJ^i_0\,,\qquad
\hat{H}=\int dx\hat{\cal H}=H-h_iQ_i\,,
\eeq
where $h_i$ are constant external fields having the interpretation of
chemical potentials.
In the limit of large fields ($h_i\to\infty$) corresponding to considering
the system at large density, due to asymptotic freedom
weak coupling perturbation theory (in $\lambda$ resp.\ $\tilde\lambda$)
is applicable.

The Legendre transform of the ground state energy density of the system
must be of the form:
\bee \label{free0}
{\cal F}(h)\equiv\df \equiv f(h)-f(0)=-h^2 F_0(h/\Lambda,Q)\,,
\eeq
where $\Lambda$ is a parameter of dimension mass.
(Eq.\ (\ref{free0}) follows simply from dimensional analysis.)
In an asymptotically free theory the result of
a perturbative computation of $F_0$
(which is of course renormalization group invariant)
is an asymptotic series in the running coupling,
$\bar\lambda(h/\Lambda)\propto[\ln(h/\Lambda)]^{-1}$,
where $\Lambda$
is the usual renormalization group invariant combination
of the (arbitrary) mass scale, $\mu$, and the renormalized
coupling, $\lambda$.

Since ${\cal F}(h)$ is a physical quantity, the results obtained in
the two models should agree.
Thus comparing them
provides a further nontrivial check on the quantum
equivalence
between the axion and the anisotropic $\sigma$-model, hence also on the
validity of the T-duality transformation at the quantum level.

Since the axion field, $\theta$, is actually an angle,
its winding number (topological charge) can be non trivial.
Therefore I present here the Legendre transformation
of the modified Hamiltonian (\ref{hamgen}) for a rather general
case including also topological charges.

Let us consider a general sigma model with torsion
\bee\label{Laggen}
{\cal L}_0=\frac{1}{2}g_{AB}\pa^\mu X^A\pa_\mu X^B+
\frac{1}{2}b_{AB}\epsilon^{\mu\nu}\pa_\mu X^A\pa_\nu X^B,
\eeq
and the following Ansatz for a set of conserved currents
\bee\label{curr}
J_\mu^i={\cal C}_A^i(X)\pa_\mu X^A+\epsilon_\mu^{\ \nu}
{\cal B}_A^i(X)\pa_\nu X^A\,,
\eeq
sufficiently general to include topological currents.
The Legendre transformation of (\ref{hamgen}) yields the
Lagrangian of the modified model which can be written as
\bee
\hat{\cal L}={\cal L}_0+h^iJ^i_0+\frac{1}{2}h^ih^j{\cal C}_A^i
{\cal C}^{A j}\,.
\eeq
In fact $\hat{\cal L}$ can be obtained by gauging ${\cal L}_0$ i.e.\
by the substitution
\bee\label{gauging}
\pa_\mu X^A\to\pa_\mu X^A +h^i\delta_{\mu 0}{\cal C}^{iA}
\eeq
when the antisymmetric field $b_{AB}$ is invariant (without compensating
gauge transformation)
under the symmetry transformation generated by the conserved currents
(\ref{curr}).

\subsection{The anisotropic PCM model}
I recall next the main results of the perturbative computation of
${\cal F}$ in the anisotropic PCM (\ref{Lagdef}) first (for more
details see Ref.\ \cite{BF}). There are two conserved Noether
charges in the model (\ref{Lagdef}), $Q_{\rm L}$, $Q_{\rm R}$,
corresponding to the U(1)$_{\rm L}\times$U(1)$_{\rm R}$
transformation
\bee
\delta G=i\epsilon_L\sigma^3G-i\epsilon_RG\sigma^3\,.
\eeq
Introducing two
chemical potentials coupled to the $Q_{\rm L}$ resp.\ $Q_{\rm R}$,
charges the Hamiltonian (\ref{hamgen}) takes the form:
\bee\label{hamgen1}
H=H_{\Sigma} - h_{\rm L}Q_{\rm L}-h_{\rm
R}Q_{\rm R}\,.
\eeq
Since the anisotropic PCM is torsionless the
Legendre transformation of (\ref{hamgen1}) can be simply found
using the gauging procedure (\ref{gauging}). To actually
calculate the ground state energy in PT one has to pass to the
Euclidean field theory formulation in which case the chemical
potentials $h_i$ can be interpreted as constant, imaginary gauge
fields. The corresponding covariant derivative $D_\mu$ is then
given by
\bee
D_2G=\partial_2G+h_L\sigma^3G-h_RG\sigma^3\,,\qquad\qquad
D_1G=\partial_1G\,. \label{cov}
\eeq
The gauged Lagrangian can be
written as:
\bee
{\cal L}={\cal L}_0+{\cal L}_1+{\cal L}_2\,,
\label{Lag}
\eeq
where ${\cal L}_1$ and ${\cal L}_2$ denote the
terms linear and quadratic in the external fields, respectively.

The computation of the free energy is based on the generating functional
\bee
e^{-\int d^nx\,f(h)}=\int{\cal D}G\,{\rm exp}\Big\{-\int d^nx\,
({\cal L}_0+{\cal L}_1+{\cal L}_2)\Big\}\,.
\label{master}
\eeq
In PT one expands Eq.\ (\ref{master}) in powers of the (bare) coupling
$\lambda_0$.
I have also written the volume element as $d^nx$, where $n=2-\epsilon$
to indicate
that dimensional regularization is employed in the perturbative
calculations.

In perturbation theory the action has to be expanded around a
(stable) solution of the classical equations of motion. This
amounts to find elements of SU(2), $G_0$, that correspond
to (local) minima of the quadratic part of the gauged Lagrangian,
\bee
{\cal L}_2=-{2h_L^2\over\lambda_0}(1+g_0z^2)-
{2h_R^2(1+g_0)\over\lambda_0}+
{4h_Lh_R(1+g_0)z\over\lambda_0}\,,
\label{Lag2}
\eeq
where
\bee
z={1\over2}{\rm Tr}\,\Big\{\sigma^3G^{-1}\sigma^3G\Big\}\,.
\eeq
The first solution, referred to as \lq bosonic', (BOS) is
given as:
\bee
G_0={1\over\sqrt{2}}
(1+i\sigma^2)\,,\quad h_R=0\,,
\label{G0BOS}
\eeq
and the other one considered here, which I call \lq fermionic' (FER) is
\bee
G_0=i\sigma^2\,.
\label{G0FER}
\eeq
(The reason for the names \lq bosonic' resp.\ \lq fermionic' will become
clear later on.)

I start with the BOS case first.
The leading, ${\cal O}(\lambda_0^{-1})$, term in perturbation theory is
given by the potential energy, ${\cal L}_2$, at its minimum, which
is immediately seen to be ($z_0=0$, $h_R=0$)
\bee
{\cal F}^{(-1)}_{\rm\scriptscriptstyle BOS}={\cal L}^{(-1)}=-{2h_L^2\over\lambda_0}
\label{FclassBOS}\,.
\eeq
The 1-loop term leads to the following integral:
\bee
{\cal F}^{(0)}_{\rm\scriptscriptstyle BOS}
={4h_L^2\over n}\int{d^np\over(2\pi)^n}\,
{(1+g_0)\,p_2^2-g_0p^2\over
(p^2)^2-4g_0h_L^2p^2+4(1+g_0)h_L^2p_2^2}\,.
\label{F0BOS}
\eeq
For details concerning the actual computation,
see Ref.\ \cite{BF}, here I just quote
the final RG improved perturbative result for
the asymptotic expansion of the free energy for large values of the
external fields:
\bee
{\cal F}_{\rm\scriptscriptstyle BOS}(h)=-{h^2\over\pi}\Bigg\{s+\Big(1-{p\over2}\Big)\ln s+
\Big(\ln2-{1\over2}\Big)+\cdots\Bigg\}\,.
\label{FexpBOS}
\eeq

In the FER case the leading (classical) term of the free energy is
\bee
{\cal F}^{(-1)}_{\rm\scriptscriptstyle FER}={\cal L}^{(-1)}=-{2(1+g_0)\over\lambda_0}\,(h_L+h_R)^2\,.
\label{FclassFER}
\eeq
Note that ${\cal F}^{(-1)}$  for the diagonal
charge depends {\sl only} on the sum $h=h_L+h_R$.
In the present case one needs to expand ${\cal F}(h)$ up to
${\cal O}(\eff^2)$ in the effective coupling, which
would at first sight necessitate a three-loop computation.
In fact one obtains all the ${\cal O}(\eff^2)$ terms from the one loop
result alone! This `mini miracle' is due to the
$(1+g)^2$ factor in front of the one-loop term
together with the fact that $(1+\bar g)^2={\cal O}(\eff^2)$, implying that
the one-loop term is already ${\cal O}(\eff^2)$.
The result turns out to be:
\bee
{\cal F}_{\rm\scriptscriptstyle FER}(h)=-{ph^2\over\pi}\Bigg\{1-{p\over4\pi}\eff+
{p^2\over32\pi^2}\eff^2+
{p\eff^2\over8\pi^2}\Bigg[\ln p+\ln\Big({\eff\over2\pi}\Big)\Bigg]+
{\cal O}(\eff^3)\Bigg\}\,.
\label{FeffFER}
\eeq
Note the non-analytic contribution, $\ln \eff$, in the last term of
(\ref{FeffFER}). The large $s$ expansion
in the FER case is finally given as
\bee
{\cal F}_{\rm\scriptscriptstyle FER}(h)=-{ph^2\over\pi}\Bigg\{1-{p\over2s}-{p^2\over4s^2}\ln s+
{p\over2s^2}\Bigg[\ln p+{p\over4}\Bigg]+\cdots\Bigg\}\,.
\label{FexpFER}
\eeq
The above form of ${\cal F}(h)$ Eq.\ (\ref{FeffFER}) explains the
name `fermionic' as its leading term corresponds to the free energy of $p$
free fermions. The ${\cal O}(1/s)$ terms can be interpreted as interaction
terms.
In the `bosonic' case the leading term in Eq.\ (\ref{FexpBOS})
is logarithmically divergent.

Finally note
that the one loop term also depends {\sl only} on the sum
$h_L+h_R$, just like
the classical one (\ref{FclassFER}). This is not an accident.
As shown in Ref.\ \cite{BF}, in the FER case the free energy does depend only
on the sum $h_L+h_R$ to {\sl all orders} in perturbation theory.
In fact it is essential that ${\cal F}(h_L,h_R)$
be a function of $h=h_L+h_R$ in order to
be able to match the perturbative result with the corresponding one
computed by the
non-perturbative Thermodynamical Bethe Ansatz
for the DIAG case in Subsection 4.2.

\subsection{The axion model}

Next I outline the computation of the corresponding ground
state energies to one loop order in the axion model (\ref{Lagax}),
starting  with the BOS case first.
With the Euler angle parametrization of $G$ (\ref{G})
the U$_{\rm L}(1)$ transformation,
$G\mapsto e^{i\kappa\tau^3}G$ of the anisotropic $\sigma$-model
(\ref{Lagdef})
acts as a simple shift, $\varphi (x)\mapsto\varphi(x)+\kappa$.
The corresponding Noether charge, $Q_{\rm L}$, and its image under
the T-duality transformation, $\tilde Q_{\rm L}$, are simply
 $$Q_{\rm L}=\int dxp_\varphi\,,\quad
 \tilde{Q}_{\rm L}=\int dx
\tilde{p}_\varphi\,,\quad
 {\rm where}\quad
 p_\varphi=\frac{\pa {\cal L}_\Sigma}{\pa\dot{\varphi}}\,,\;
\tilde{p}_\varphi=\frac{\pa {\cal L}}{\pa\dot{\varphi}}\,,
$$
since the canonical transformation
implementing the T-duality mapping (\ref{cano}) effects only
$p_\alpha$, $\chi^\prime$,
$\alpha^\prime $ and $p_\chi$,
leaving the other fields,
$\varphi $, $\vartheta $, $p_\varphi$, $p_\vartheta$, unchanged.

Since in the BOS case the $b_{AB}$ field in Eq.\ (\ref{Lagax}) is invariant,
one can simply `gauge' the Lagrangian of the
axion model in an external ($h_{\rm L}$) field (see Eq.\ (\ref{gauging}).
The classical ground state is found to be
 $\varphi\equiv\chi\equiv 0$, $\vartheta\equiv
{\pi}/2$.
(The corresponding solution of the anisotropic $\sigma$-model
is given by $\varphi\equiv\alpha\equiv 0$, $\vartheta\equiv
{\pi}/2$.)

Expanding the (Euclidean) Lagrangian (after suitable rescalings,
etc.) one obtains
\bee\label{kozlag}
\overline
 {\cal L}=-\frac{2h^2_{\rm L}}{\tilde{\lambda}_0}
+ \frac{1}{2}m{\cal
M}m^T+{\tt o}(\tilde{\lambda}),
\eeq
where
\bee\label{du0}
{\cal M}=\pmatrix{-\pa^2+4h_{\rm L}^2&0&2h_{\rm
L}\sqrt{1+\tilde{g}_0}\, \epsilon_{\mu 2}\pa_\mu\cr 0&-\pa^2&0\cr
-2h_{\rm L}\sqrt{1+\tilde{g}_0}\,\epsilon_{\mu 2}\pa_\mu&0&-\pa^2\cr},
\eeq
and
 $m =(\vartheta, \varphi ,\chi)$. ($\tilde{\lambda}_0$, $\tilde{g}_0$
 denote the bare coupling and parameter of the axion model).
 In Eq.~(\ref{du0}) the $\epsilon$
tensor has been explicitly kept, as it requires a careful definition in
$n=2-\epsilon$ dimensions  used to regularize the momentum
integrals. It is convenient to adopt the definition of Ref.\ \cite{Osb},
where this
antisymmetric tensor corresponds to an almost complex structure:
$\epsilon_{\mu\nu}=-\epsilon_{\nu\mu}$,
$\epsilon_{\mu\nu}\epsilon_{\mu\sigma}=\delta_{\nu\sigma}$. The
one loop quantum corrections to the classical ground state (the
first term in Eq.\ (\ref{kozlag})) require the calculation of a
functional determinant, leading to
\bee\label{det}
{\cal F}(h)=\frac{4h_{\rm L}^2}{n}\int\frac{d^np}{(2\pi )^n}
\frac{\tilde{p}_1^2-\tilde{g}_0\tilde{p}_2^2}{{\tilde{p}}^4+
4h_{\rm L}^2(\tilde{p}_1^2- \tilde{g}_0\tilde{p}_2^2)}\,,\quad
\tilde{p}_\mu=\epsilon_{\mu\nu}p_\nu\,.
\eeq
To evaluate
(\ref{det}) one should apply the modified dimensional regularization of
Ref.\ \cite{BF} as $\tilde{p}_2=\epsilon_{2\nu}p_\nu$ plays
a distinguished r\^ole and it is kept as a one dimensional variable.
In fact for our purposes it is sufficient to calculate the {\sl
difference} ${\cal F}(h)-{\cal F}_{\Sigma}(h)$, where ${\cal
F}_{\Sigma}(h)$ is the corresponding determinant in the anisotropic
$\sigma$-model (Eq.\ (3.12) in Ref.\ \cite{BF}). Since both
${\cal F}(h)$ and ${\cal F}_{\Sigma}(h)$ are already the first
quantum corrections to the classical expressions one may set
$\tilde{g}=g$ (and make no distinction between bare and
renormalized $g$'s) when computing their difference to lowest
order and one ends up with
\bee\label{F0dif}
{\cal F}(h)-{\cal F}_{\Sigma}(h)= \frac{(2h_{\rm
L})^n}{n}(1+g)\int\frac{d^nq}{(2\pi )^n}
\frac{(q_1^2-q_2^2)q^4}{N_1N_2}=\frac{(2h_{\rm L})^n}{n}(1+g)w(g),
\eeq
where $N_1=q^4+q_1^2-gq_2^2$, $N_2=q^4+q_2^2-gq_1^2$.
Although the integrand yielding $w(g)$ is antisymmetric under
$q_1\leftrightarrow q_2$, the integral is divergent by power
counting for $n=2$, i.e.\ it must be computed in $n=2-\epsilon$
dimensions. Its derivative, $w^\prime (g)$, is, however, {\sl
convergent} by power counting and it has also an antisymmetric
integrand, therefore this latter may be evaluated in $n=2$
dimensions giving $w^\prime (g)\equiv 0$. Then to compute $w(g)$
one may choose e.g.\ the point $g=-1$:
\bee\label{wme}
w(-1)=\int\frac{d^{n}q}{(2\pi )^n}
\frac{q_1^2-q_2^2}{(q^2+1)^2}=\frac{n-1-1}{n}
\int\frac{d^nq}{(2\pi )^n} \frac{q^2}{(q^2+1)^2}=-\frac{1}{4\pi},
\eeq
 where writing the second equality, it has been used that $q_1$ is
an $n-1$-dimensional variable, while $q_2$ is $1$-dimensional.
From (\ref{wme}) one finds that after taking into account the
change of the renormalization scheme (\ref{psicoupl}), in PT the
free energy densities of the two models (\ref{Lagax}) and
(\ref{Lagdef}) do indeed coincide for the BOS case.

I omit the details of the calculations
in the FER case (see Ref.\ \cite{BFP} for details), suffice it to say that
this free energy is also in perfect agreement with the corresponding
result in the anisotropic PCM model.

I would like to point out here that
the canonical transformation connecting ${\cal L}_\Sigma$ and
${\cal L}$, maps $Q_{\rm R}$ to a purely {\sl topological
charge} $\tilde{Q}_{\rm R}$ of the axion model, which is completely
different from the Noether charge of the `manifest' SO(2)$_\theta$
symmetry of the Lagrangian (\ref{Lagax}).
To illustrate this I quote the value of the classical free
energy density corresponding to
the Noether charge of the SO(2)$_\theta$ symmetry:
$\hat{H}\vert_{\rm min}^{(\theta)}=-{2h_{\rm R}^2}/{\tilde{\lambda}}$,
very different indeed from Eq.\ (\ref{FclassFER}) with $h_{\rm L}=0$.

In conclusion the free energy densities
fully agree for the BOS and FER cases in both models
up to first order in  RG improved PT, providing a rather convincing
evidence for the validity of T-duality at the quantum level.

\setcounter{section}{3} \setcounter{equation}{0}
\section{Quantum integrability}

It is by now generally accepted that the anisotropic PCM model
(\ref{Lagdef}) is integrable at the quantum
level \cite{Kirillov,BF,Fateev} in the sense that there is no
particle production. Assuming that the spectrum contains two
massive doublets (kinks) their scattering is described by the
tensor product of an SU(2)$\times$U(1) symmetric solution of the
bootstrap $S$-matrix equations:
\bee
S(\theta)=S^{(\infty)}(\theta)\otimes S^{(p)}(\theta)\,,
\label{Smatrix}
\eeq
where $S^{(p)}(\theta)$ denotes the
Sine-Gordon (SG) $S$-matrix. Let me point out here that denoting
the parameter in the $S$ matrix in the same way as the RG
invariant quantity in Eq.\ (\ref{invp}),($p$), is intentional. As
will become clear from the results later on, one should actually
{\sl identify} them. (In fact this identification is seen to be
consistent at least up to two loops in PT.)

In the limit $p\to\infty$ (corresponding to $\bar g\equiv-1$)
the symmetry of the $S$-matrix (\ref{Smatrix})
increases to SU(2)$\times$SU(2).
For the range of the parameter $0<p<1$ in addition to the kinks
bound states (breathers) also appear
in the spectrum, transforming as $3+1$ under SU(2).

In this section I shall explore some consequences of
assuming the validity of the duality transformation at the
{\sl full quantum level} between the two theories.
This immediately implies the absence of particle production in the axion
model (\ref{Lagax}) too, and that its two particle $S$-matrix
is given by Eq.\ (\ref{Smatrix}). Note that the
global symmetries of the axion model,
SO(3)$\times$SO(2)$_\theta$, {\sl cannot be identified} with those of the
$S$ matrix (\ref{Smatrix}). As it will be convincingly shown
the U(1) symmetry of the $S$ matrix (\ref{Smatrix}) corresponds to
a hidden `topological' U(1) of the axion model,
the dual of the U(1)$_{\rm R}$ Noether symmetry
of the anisotropic PCM under the T-duality transformation.

The following (rather heuristic) consideration might be useful to
give some insight into the connection between the
SG and the axion model.
Integrating out the O(3) fields, $n^a$, in some
generating functional of the theory (\ref{Lagax})
one obtains a non-vanishing effective potential for the $\theta$
field. Since $\theta$ is $2\pi$-periodic the effective
potential must be also periodic. The effective theory of the
$\theta$ field is therefore expected to show some features
similar to the Sine-Gordon model, with a periodic potential and
corresponding topological current $K_\mu=\epsilon_{\mu\nu}
\partial^\nu\theta/2\pi$.
(Note that $\theta$ being an angular variable makes it difficult
to integrate it out in the functional integral in spite of ${\cal
L}$ being only quadratic in $\theta$.)

\subsection{Particle production in the O(3) model with a $\theta$-term}

At first sight the proposed quantum integrability of the axion model
seems to be rather questionable at least, as
it is generally believed that the
O(3) $\sigma$-model with a {\sl constant} $\theta$-term
is not quantum integrable,
except for the special value $\theta=\pi$ \cite{SMpi}
(despite the fact that the $\theta$-term, being a total derivative,
does not change the classical physics of the model).

In the framework of the form-factor bootstrap
approach one can actually show that the $\theta$ term
mediates particle production, indeed \cite{BFP}.
To illustrate this important fact, let us express
the $2\to3$ particle production amplitude
to lowest order in $\theta$ as:
\bea
&&\langle p,b;p^\prime,b^\prime;p^{\prime\prime},
b^{\prime\prime}\vert q,a;q^\prime,a^\prime\rangle_{(\theta)}=
(2\pi)^2\,i\,\theta\,\delta^{(2)}
(p+p^\prime+p^{\prime\prime}-q-q^\prime)\nonumber\\
&&\qquad\qquad \cdot\,
\langle p,b;p^\prime,b^\prime;p^{\prime\prime},b^{\prime\prime}\vert
T(0)\vert q,a;q^\prime,a^\prime\rangle_{(0)}+
{\cal O}(\theta^2)\,,
\label{amplitude}
\eea
where in the first line the amplitude is in the
O(3) model with a $\theta$-term, while in the second line the
matrix element of the topological charge density operator
$T$ is to be
calculated in the original O(3) $\sigma$-model (with $\theta=0$).
In other words one computes the $2\to3$ particle production amplitude
to first order in $\theta$.

For the sake of simplicity let us consider the following special
kinematical configuration: the incoming particles have momenta $q_1=Q$
and $q_1^\prime=-Q$, whereas the produced (outgoing) three particles
have momenta $p_1=Q^\prime$, $p_1^\prime=0$ and $p_1^{\prime\prime}=
-Q^\prime$ respectively. Here $Q^\prime$ can easily be expressed in
terms of $Q$ and the kink mass $M$ using energy conservation. For
large $Q$, using the results of Ref.\ \cite{BN}, one finds
\bee
\langle p,b;p^\prime,b^\prime;p^{\prime\prime},b^{\prime\prime}\vert
T(0)\vert q,a;q^\prime,a^\prime\rangle_{(0)}\approx
{\pi}^{5\over2}\frac{Q^2}{\ln^3Q/M}\left(\epsilon^{a^\prime ba}\delta^{
b^\prime b^{\prime\prime}}-
\epsilon^{b^{\prime\prime} ba}\delta^{
b^\prime a^\prime}\right)\,.
\label{asy}
\eeq
Eq.\ (\ref{asy}) shows that already to first order in $\theta$, the
$2\to3$ particle production amplitude is different from zero.
Thus at least for small values of $\theta$, the introduction of this
term
destroys quantum integrability of the O(3) $\sigma$-model, indeed.
To the best of my knowledge this is the first quantitative evidence for
the quantum non-integrability of the O(3) $\sigma$-model with a
{\sl constant} $\theta$-term.

\subsection{Free energy from the Thermodynamical Bethe Ansatz}

The free energy of a system of particles of mass $m$ is of the form:
\bee \label{free1}
{\cal F}(h)=-h^2 F_0(h/m,Q)\,.
\eeq
The Thermodynamical Bethe Ansatz (TBA) method
based on the $S$--matrix of the particles
leads to a set of integral equations for the free energy
whose solution can be found in many cases
as an asymptotic series expansion in $h/m\gg1$.
Since ${\cal F}$ is a physical quantity, the results obtained by TBA
method and by perturbation theory should agree.
Thus by comparing the asymptotic series of $F_0(h/m,Q)$ in $h/m$ with
the one of $F_0(h/\Lambda,Q)$ in $h/\Lambda$ obtained in PT
 provides a rather stringent consistency check on the
$S$-matrix (and also on the self-consistency of the hypothesis used in the
course of the calculation). Moreover
one obtains the exact $m/\Lambda$ ratio, a rather important
non-perturbative parameter of the theory which
can be measured e.g.\ by lattice simulations.

For definiteness we choose \hbox{$h_{\rm L}, h_{\rm R}\geq0$}
and assume $p>1$, implying that there are two doublets in the
spectrum. The charges of the particles are normalized as:
\bee\label{charges}
(1,1)\,,\quad(-1,1)\,,\quad(1,-1)\quad(-1,-1)\,.
\eeq
One can distinguish between the three possible types
of finite density ground states depending on $h_{\rm L}, h_{\rm R}$ as:
\begin{itemize}
\item[1.] $h_{\rm L}\,, h_{\rm R}>0\,,\!\qquad\qquad$  (DIAG)
\item[2.] $h_{\rm L}=0$\,, $h_{\rm R}=h>0\,,$ (RIGHT)
\item[3.] $h_{\rm L}=h>0$\,, $h_{\rm R}=0\,.$ (LEFT)
\end{itemize}

\vskip6pt
\noindent
$\bullet$ The DIAG case.\\
As the system (\ref{hamgen1}) is considered
with $h_{\rm L}/m$, $h_{\rm R}/m\gg1$, it is clear that particles of
charge (1,1) condense into the vacuum. It is less clear what other
kind of particles (necessarily with a smaller charge/mass ratio)
will appear in the vacuum state. I shall assume that the vacuum
consist of {\sl only} particles of charge (1,1). This seemingly
radical assumption has apparently worked in all analogous examples
studied so far, and it greatly simplifies the solution of TBA
equations.

The scattering phase of the (1,1) type particle, $\delta(\theta)$,
can be easily found from the $S$-matrix:
\bee\label{diagphase}
\delta(\theta)=\delta_\infty(\theta)+\delta_p(\theta)\,.
\eeq
When the ground state is assumed to contain {\sl only} particles of
charge (1,1) the calculation of the free energy using the TBA
method reduces to the solution of a single integral equation
(\ref{inteq}) whose kernel is given by the logarithmic derivative
of the relevant $S$-matrix element,
\bee K(\theta)={1\over2\pi i}{d\over d\theta} \ln S(\theta)\,.
\eeq
The integral equation in
this case is given as follows:
\bee\label{inteq}
\epsilon(\theta)-\int_{-B}^{B}d\theta'K(\theta-\theta')
\epsilon(\theta')=h-m\cosh\theta\,,
\eeq
where $h=h_{\rm L}+h_{\rm R}$, together with the boundary condition
$\epsilon(\pm B)=0$. In
terms of the solution of Eq.~(\ref{inteq}) the free energy is
\bee
\delta f(h)=-{m\over2\pi}\int_{-B}^{B}d\theta\cosh(\theta)\epsilon(\theta)\,.
\eeq
The asymptotic series for the free energy up to $O(\ln t/t^3)$
terms reads as:
 \bee\label{ferfreen}
\delta{f}_{\rm D}(h)=-{ h^2 \over\pi}\,p \left\lb 1-{p\over2t}- {p^2\over4}
{\ln t\over t^2}+
{A_{\rm D} \over t^2}+O({\ln t \over t^3}) \right\rb\,,
\eeq
 where
$t=\ln(h/m)$ and $A_{\rm D}$ is a constant which determines the
$m/\Lambda$ ratio:
\bee \label{atild} A_{\rm
D}={p\over2}\left[\ln\Gamma(1+{p\over2})+{p\over2}({3\over2}-\ln2)
-1+3\ln2 -\ln\pi\right]\,.
\eeq
 Let me remark here that from the
classical term in Eq.\ (\ref{ferfreen}), $-h^2p/\pi$, one can
immediately read off the level of an underlying ultraviolet (UV)
current algebra, $k$, as $k=p$ \cite{fendley}.

\vskip6pt
\noindent
$\bullet$ The RIGHT case.\\
In analogy to the assumption made in the diagonal case one would now
expect the ground state to consist of a mixture of
particles of the same charge with respect to U(1)$_{\rm R}$,
i.e.\ those of charge (1,1) and (-1,1) with equal
densities. In Ref.\ \cite{Fateev} the result of the
diagonalisation of the pertinent coupled TBA system has been given and
one finds (somewhat surprisingly)
 that
\bee
\delta f_{\rm R}(h)=\delta f_{\rm D}(h)\,,
\eeq
where $h=h_{\rm R}$.
So the free
energies are precisely the same as functions of the effective chemical
potentials in these two apparently rather different cases.

\vskip6pt
\noindent
$\bullet$ The LEFT case.\\
Analogously to the RIGHT case
the ground state is expected to consist of an
equal density mixture of
particles of the same charge, this time with respect to U(1)$_{\rm L}$,
i.e.\ those of charge (1,1) and (1,-1).
Here I just quote \cite{BF} the asymptotic series of the free energy
density:
\bee \label{bosfreen}
\delta{f}_{\rm L}(h)=-{ h^2 \over\pi}\,\,\left(t+(1-{p\over2})\ln t +
A_{\rm L}+\ldots\right)\,,
\eeq
where the constant $A_{\rm L}$,
determining the ratio $m/\Lambda$ is:
\bee \label{a}
A_{\rm L}=\left(3-{p\over2}\right)\ln2+{p-3\over2}-
\ln\pi+\ln\Gamma\left({p\over2}\right)\,.
\eeq
\vskip6pt
\noindent
$\bullet$ The LEFT case for $p<1$.\\
In the previous subsections it has been assumed that the parameter
$p>1$. In order to clarify what happens when $p$ becomes smaller
than one, I recall that then bound states (breathers) appear in
the spectrum (for $p<1$ one enters simply the attractive regime).
The SG mass spectrum is given by the well known formula:
\bee\label{SGspectrum}
m_r=2m\sin {\pi pr\over2}\,,\qquad
r=1,2\ldots <{1\over p}\,,
\eeq
where $m$ denotes the mass of the
SG kinks. (Note that $m$ also gives the mass of the SU(2)
doublets.) The appearance of these new particles implies that as
soon as $p$ becomes smaller than $1$ the vacuum changes as then it
becomes energetically favorable for the $r=1$ charge (2,0)
breather to condense. This charge (2,0) breather can be
interpreted as the bound state of the (1,1) and (1,-1) particles
(kinks). In fact as this particle has the highest charge/mass
ratio, one can again assume (as in the DIAG case) that the true
vacuum consists {\sl only} of the condensate of the $r=1$ charge
(2,0) breather. Calculating the breather-breather phase shift by
the bootstrap-fusion method \cite{Karowski} and solving the
integral equation (\ref{inteq}) yields the free energy density of
the breather condensate, $\delta{f}_{\rm B}(h_{\rm B},m_1)$, where
$m_1=2m\sin {\pi p\over2}$ denotes the breather mass. One finds
that $\delta{f}_{\rm B}(h_{\rm B},m_1)$ becomes {\sl identical} to
the free energy of the charge $(1,\pm1)$ kink mixture,
$\delta{f}_{\rm L}(h,m)$ analytically continued for $p<1$. This is
very reassuring as the result for $\delta{f}_{\rm L}(h,\Lambda)$
from perturbation theory turns out to be {\sl continuous} for the
full range $0\leq p<\infty$, which is certainly to be expected.

\subsection{The $m/\Lambda_\MSb$ ratio}
Having calculated the asymptotic form of the free energy both in
perturbation theory and with the TBA method,
by comparing them (with $t=s-\ln(m/\Lambda_{\overline{\rm MS}})$)
one obtains the relation between the mass of the
doublet particles, $m$, and $\Lambda_{\overline{\rm MS}}$.
In both the BOS (LEFT) resp.\ FER (RIGHT,DIAG) cases the comparison of
Eqs.\ (\ref{bosfreen})
and (\ref{FexpBOS}) resp.\ (\ref{ferfreen}) and (\ref{FexpFER}) leads
to the result
\bee
{m\over\Lambda_{{\overline{{\rm MS}}}}}=2^{3-{p\over2}}\,
e^{{p\over2}-1}\,
{\Gamma\big(1+{p/2}\big)\over\pi p}\,.
\label{MperLambda}
\eeq
I would like to emphasize that it is already a very nontrivial
check on the overall
consistency of the assumptions that {\sl all} of the expansion
coefficients, for both the BOS and the FER cases, agree.

To illustrate how nontrivial
this consistency is, let me point out the following
(apparent) paradox.
In Eq.\ (\ref{ferfreen}) the ratio of the
coefficients of the $\ln t/t^2$ term,
 $-p^2/4$, and the $-1/t$ term,
$p/2$, is $r_{\scriptscriptstyle\rm D}=-p/2$, while in Eq.\ (\ref{bosfreen})
the ratio of the coefficients of the $\ln t$ term, $1-p/2$,
and that of the $t$ term  is $r_{\scriptscriptstyle\rm L}=1-p/2$.
This is quite remarkable, since on very general grounds
in a theory with a coupling, $\lambda$, the perturbative result
for a fermionic type free energy is expected to be of the form
\bee
\delta{f}_{\rm Fer}(h)=-{ h^2 f_0\over\pi}\,\,\left(1-f_1{
\bar\lambda(h)\over2\pi}+
{\cal O}(\bar\lambda^2)\right)\,,
\eeq
while in the {\sl same} theory for a bosonic case it is expected to
be given by
\bee
\delta{f}_{\rm Bos}(h)=-{ h^2 b_0\over\pi}\,\,\left({1\over\bar\lambda(h)}+
{\rm const.}+ {\cal O}(\bar\lambda)\right)\,,
\eeq
where $f_0$, $f_1$, $b_0$ are constants and $\bar\lambda(h)$ is the running
coupling.
Since from the renormalization group
\bee
{1\over\bar\lambda(h)}=\beta_0\ln{h\over\Lambda}+
{\beta_1\over\beta_0}\ln\ln{h\over\Lambda}+{\cal O}({1\over h/\Lambda})\,,
\eeq
one expects $r_{\rm Fer}=r_{\rm Bos}=\beta_1/\beta_0^2$.
As in the present case
$r_{\scriptscriptstyle\rm D}\ne r_{\scriptscriptstyle\rm L}$, it seems
impossible to match Eq.\ (\ref{ferfreen})
and Eq.\ (\ref{bosfreen}) {\sl simultaneously} with the
corresponding perturbative expansion. The resolution of this
paradox lies in the presence of
the non-analytic contribution, $\ln \eff$, in the last term of
Eq.\ (\ref{FeffFER}).
This non-analytic term in $\eff$
explains why the coefficient of the $\ln s$ term differs
for the BOS and the FER cases. Such non-analytic
terms can only occur in perturbation theory in the presence of several
couplings.

A further, very stringent
consistency check is that the $m/\Lambda$ ratio obtained in the BOS
case is exactly the same as the one obtained in the FER case.
Finally by comparing the corresponding free energies
of the breather condensate one obtains
the $m/\Lambda$ ratio in Eq.\ (\ref{MperLambda}).
This is clearly an additional check on our result
and on the mutual consistency of the hypothesises made in the course of
the calculation.

\subsection{The $p\to 0$ limit}
The $p\to0$ limit (i.e.\ the running
deformation parameter, $\bar g(h)\to-1$),
is expected to give the
O(3) $\sigma$-model (and a decoupled free field) \cite{Wiegmann}.
In fact from Eq.\ (\ref{SGspectrum})
one sees that $m/m_1\to\infty$ as $p\to 0$ i.e.\ the kinks
`disappear' from the spectrum. Furthermore the higher breathers
with $r>1$ also `disappear' from the spectrum, but for a different
reason.
The binding energy of the $r>1$ breathers
(which can be considered as bound states of the $r=1$ ones)
tends to zero giving rise to zero energy bound states.
Therefore the limiting spectrum consists of 4 massive particles,
transforming as $(3_{\rm L}+1_{\rm L}, 1_{\rm R})$ under
SU(2)$_{\rm L}\times$U(1)$_{\rm R}$.
Our results
lend some additional support to this expectation, as the
$m_1/\Lambda_{{\overline{{\rm MS}}}}=8/e$ ratio in the O(3)
$\sigma$-model calculated previously \cite{history}
(where now $m_1$ stays for the
mass of the triplet in the O(3) model) is beautifully reproduced in the
$p\to0$ limit.
This can be immediately seen from Eq.\ (\ref{MperLambda}) using the fact
that in the $p\to0$ limit $m_1\to \pi pm$.

\section{Conclusions}

In conclusion one can say that there is complete consistency
between the TBA and the perturbative calculations in both models,
providing good evidence for the validity of
the proposed $S$ matrix (\ref{Smatrix}) for them.
Since the effective
coupling (\ref{betaeff1}) is identical in the two models
the $m/\Lambda_\MSb$ ratio Eq.\ (\ref{MperLambda}) stays also the same.
It might be worth emphasizing once again that these results
provide also nonperturbative evidence for the validity of
quantum T-duality. In particular one has some quantitative
evidence for the identification between the U(1)$_{\rm R}$ symmetry
of the proposed $S$ matrix
and the topological current of the axion model
(corresponding to the
canonical image of the U(1)$_{\rm R}$ Noether current of
the anisotropic PCM and
being quite distinct from the manifest SO(2)$_\theta$ symmetry).

The above findings provide good motivation to study both models
by lattice Monte-Carlo simulations.
Measuring the $m/\Lambda_L$ ratios ($\Lambda_L$ denoting the
$\Lambda$-parameter
of the lattice regularized theory)
would provide us with a completely non-perturbative
way of testing quantum T-duality.

\bigskip

\noindent{\bf Acknowledgements}\\
I would like to thank the organizers of the Johns Hopkins
Workshop 2000, in particular Zal\'an and Laci,
for providing me the opportunity to present the above results
and Max Niedermaier for a careful reading of the manuscript.


\begin{thebibliography}{99}
%
\bibitem{class}
J.~Balog, P.~Forg\'{a}cs, Z.~Horv\'{a}th and L.~Palla,
Phys.~Lett.~{\bf 324B} (1994) 403.
%
\bibitem{Busch} T.H.~Buscher, Phys. Lett. {\bf B201} (1988) 466,
ibid. {\bf B194} (1987) 59.
%
\bibitem{Alvarez}E.~Alvarez, L.~Alvarez-Gaum\'{e}, Y.~Lozano,
Phys. Lett.  {\bf B336} (1994) 183.
%
\bibitem{Cherednik}
I.V.~Cherednik, Theor.~Math.~Phys.~{\bf 47} (1981) 422.
%
\bibitem{Kirillov}
A.~Kirillov, N.Yu.~Reshetikhin in Proc.\ of Proceedings of the Paris-Meudon
Colloquium
String Theory, etc. (1986), eds.\ N.~Sanchez, H.~de Vega,
(World Scientific, Singapure).
%
\bibitem{bfhp}J.~Balog, P.~Forg\'{a}cs, Z.~Horv\'{a}th, L.~Palla,
  Nucl.~Phys.~B (Proc. Suppl.)
{\bf 49} (1996) 16. ({\tt hep-th/9601091})
%
\bibitem{alvtop} E.~Alvarez, L.~Alvarez-Gaum\'{e}, J.L.F.~Barb\'{o}n and
Y.~Lozano, Nucl. Phys. {\bf B415} (1994) 71.
%
\bibitem{Byt}
A.G.~Bytsko, J. Math. Sciences 85 (1997) 1619.
({\tt hep-th/9403101})
%
\bibitem{BF}
J.~Balog and P.~Forg\'{a}cs, Nucl.~Phys.~{\bf B570} (2000) 655.
%
\bibitem{Osb}H.~Osborn, Ann. Phys. {\bf 200} (1990) 1.
%
\bibitem{BFP}
J. Balog, P. Forg\'{a}cs, L. Palla,
Phys.~Lett.~{\bf B484} (2000) 367
%
\bibitem{Fateev}
V.A.~Fateev, Nucl.~Phys.~{\bf B473}[FS] (1996) 509.
%
\bibitem{SMpi}
I.~Affleck Field theory methods and critical phenomena, in
Fields, strings and critical phenomena, ed.\ E.~Br\'{e}zin and
J.~Zinn-Justin (North Holland, Amsterdam,1990);\\
V.A.~Fateev and Al.B.~Zamolodchikov,
Phys. Lett. {\bf 271B} (1991) 91;\\
A.B.~Zamolodchikov and Al.B.~Zamolodchikov,
Nucl.~Phys.~{\bf B379} (1992) 602.
%
\bibitem{BN}
J.~Balog and M.~Niedermaier,
Nucl.~Phys.~{\bf B500} (1997) 421.
%
\bibitem{fendley}
P.~Fendley and K.~Intriligator,
Phys. Lett. 319B (1993) 132;
%
\bibitem{Karowski}
M.~Karowski,
Nucl.~Phys.~B153 (1979) 244.
%
\bibitem{Wiegmann}
P.B.~Wiegmann,
Phys. Lett. 152B (1985) 209.
%
\bibitem{history}
A.~Polyakov, P.B.~Wiegmann, Phys. Lett. {\bf 131B} (1984) 121,\\
G.~Japaridze, A.~Nersesyan and P.~Wiegmann, Nucl. Phys. {\bf B230}
(1984) 511,\\
P.~Hasenfratz, M.~Maggiore and F.~Niedermayer,
Phys.~Lett.~{\bf 245B} (1990) 522.
%
\end{thebibliography}
\end{document}